\documentclass[aps,prl,letterpaper,twocolumn,showpacs,superscriptaddress,floatfix,longbibliography]{revtex4}
\usepackage[english]{babel}
\usepackage{amsmath}
\usepackage{color}
\usepackage{epsfig}
\usepackage{graphicx}
\usepackage{bm}
\usepackage{times,amsmath,amssymb}
\usepackage{subfigure}
\usepackage{amsfonts}
\usepackage{amssymb}
\usepackage{color}
\usepackage{mathtools}
\usepackage{empheq}
\usepackage{dsfont}

\global\long\def\bra#1{\langle #1 |}
\global\long\def\ket#1{| #1 \rangle }

\global\long\def\al{\alpha} 
\global\long\def\ga{\gamma} 
\global\long\def\De{\Delta} \global\long\def\Ga{\Gamma}
\global\long\def\th{\theta}
 
\global\long\def\si{\sigma} \global\long\def\vfi{\varphi}
\global\long\def\ro{\rho}

\global\long\def\bege{\begin{equation}}
\global\long\def\ende{\end{equation}}
\global\long\def\begal{\begin{align}}
\global\long\def\endal{\end{align}}
\begin{document}
\title{Switching pure states of the dissipative Heisenberg $XXZ$ chain by local  magnetic fields}

\author{Vladislav Popkov}
\affiliation{Department of Physics, Bergische Universit\"at Wuppertal,  42097 Wuppertal Germany}
\affiliation{Department of Physics,  University of Ljubljana,
Jadranska 19, SI-1000 Ljubljana, Slovenia}
\author{Mario Salerno}
\affiliation{Dipartimento di Fisica ``E.R. Caianiello'', and Istituto Nazionale di Fisica Nucleare - Gruppo Collegato di Salerno, Universit\'a di Salerno, Via Giovanni
Paolo II, 84084 Fisciano (SA), Italy}
\begin{abstract}
The effects of a local magnetic field  on nonequilibrium stationary states (NESS) of the open quantum $XXZ$ spin chain are investigated  with a Lindblad master equation approach in the limit of strong dissipation. The local magnetic field is applied to a single bulk spin of the chain while the ends are kept at fixed polarizations by dissipation. We show that suitable changes of the local magnetic field permit to invert the spin current by switching pure NESS with opposite chiralities, while preserving the state purity and achieving optimal transport.
\end{abstract}
\maketitle
{\it Introduction.}
Much efforts are presently devoted to the study of quantum  properties, such as entanglement and purity, that allow to select within the huge Hilbert space of a quantum many-body system, states that are useful for quantum information and for the development of new quantum devices. As it is well known, entangled states are fundamental for quantum protocols involving key distribution, error corrections, teleportation, etc., while pure  states play a crucial role for security and efficiency of these protocols \cite{Nielsen}. In the real world, however, a quantum system is always in  contact with some  environment which leads to decoherence and therefore to the loss of the above properties quickly in time. This implies that quantum states of realistic many-body systems are rarely pure.

Recently it has became clear that it is possible to stabilize quantum states by means of dissipation~
\cite{2016Zanardi,PhysRevA.87.033802,Cormick2013NJP,PhysRevLett.110.120402,LinNature2013Bell,
PhysRevA.88.023849,Yi2012,1367-2630-14-6-063014,TicozziViola2012,Cirac2011PRL107,2016Kondo}.
This approach has many practical advantages with respect to the creation of the same states via coherent dynamics, since the quantum states produced by means of dissipation  are intrinsically nonequilibrium stationary states (NESS) not reachable within Gibbs ensemble and much more resistant against decoherence. 

On the other hand, the management of NESS is far from being trivial and typically requires the use of sophisticated dissipative long range mechanisms~\cite{ZollerNature2008}. In particular, to change a dissipatively produced pure state into another pure state with qualitatively different physical properties, nontrivial changes of either the coherent part of the dynamics, or of the dissipators, or of both, are required~\cite{Cirac-Nature2009}. All these operations involve structural changes that are not easy to implement in real experiments.
For practical applications it would be therefore convenient if the management of dissipatively produced states could be realized by means of external fields, without making any structural change to the system. In particular, if two pure  NESS carrying opposite currents could be switched under this type of management, this would allow to realize a  quantum switch that inverts the current simply by means of a local field.
\begin{figure}[ptbh]
\vskip -2.4cm
\begin{center}
{\includegraphics[width=0.42\textwidth]{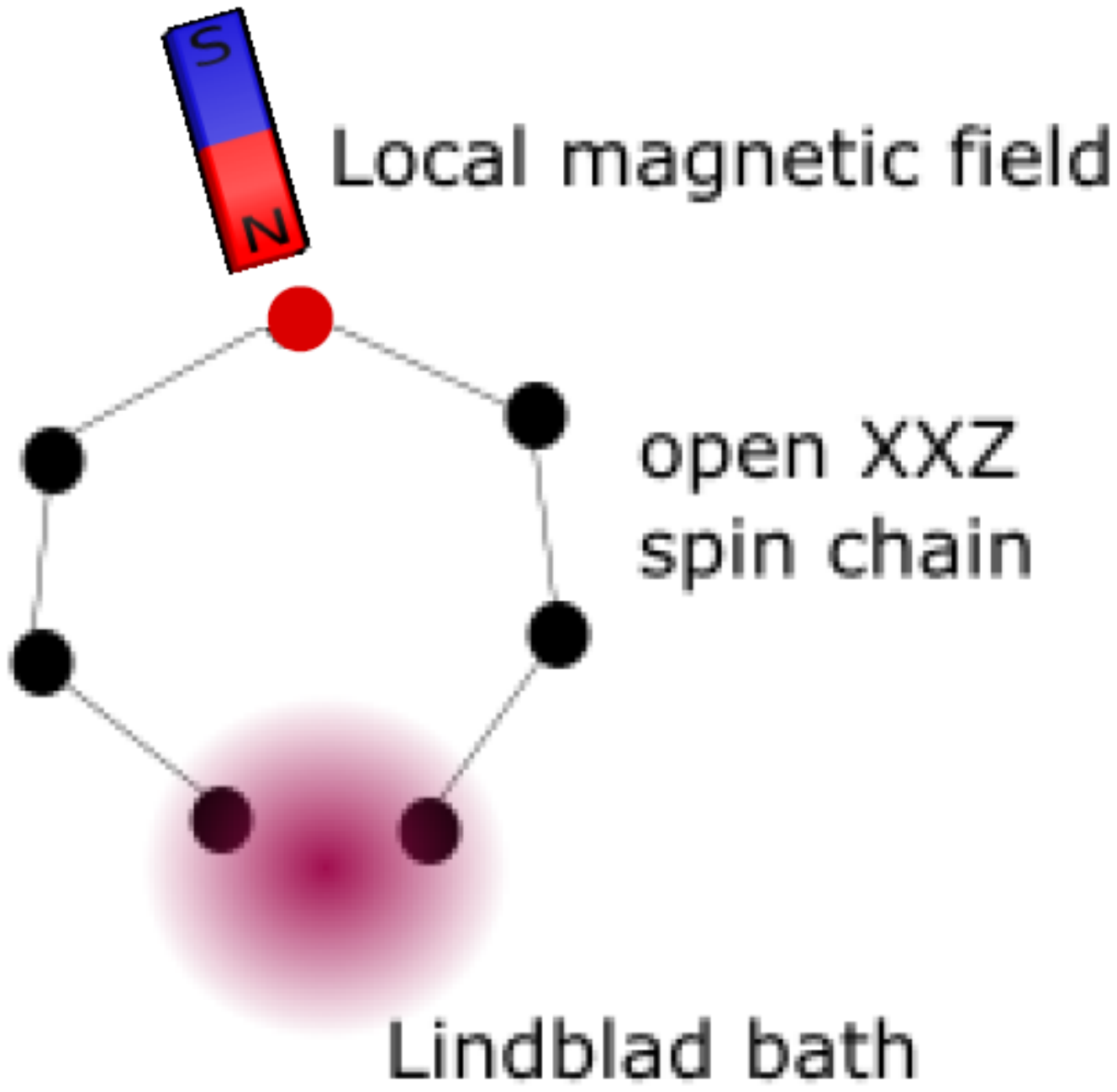}}
\end{center}
\vskip -3.6cm
\caption{
Schematic setup for a quantum switch. A spin chain strongly coupled to a dissipative magnetization bath at the ends is controlled by a local magnetic field applied to a single interior spins of the chain.
} \label{fig-Switch}
\end{figure}

The purpose of this Letter is to show how such a quantum switch can actually be made.
For this we investigate the NESS of an open quantum $XXZ$ spin chain in the limit of strong dissipation and in the presence of a local magnetic field applied to a single spin in the interior of the chain, while the ends are kept at fixed polarizations by dissipation (see Fig.~\ref{fig-Switch} for a schematic setup).
Quite remarkably, we find that suitable changes of the local field permit to switch between pure NESS with opposite chiralities and  to invert the spin current while preserving the purity of the states and achieving optimal transport. For possible practical implementations of our theoretical concept, we provide explicit estimates of the relaxation times and fidelities of the NESS involved in the switch  for the specific case of a chain of five spins.

We remark that in absence of magnetic fields the optimal transport (maximal magnetization current) in open $XXZ$ chains is always achieved in the proximity of pure states~\cite{SP-2013} and in the limit of infinitely large dissipative couplings, i.e. the so called Zeno limit~\cite{SudarshanZeno1977,KoshinoPhysRepZeno2015}, it is possible to  target pure NESS that carry high ballistic currents and remain pure on a very long time scale ~\cite{2016PopkovPresilla,2017PRAPopkovPresillaSchmidt}. Dissipative $XXZ$ spin chains have also been recently proposed as sensitive magnonic logical gates~\cite{Landi-mlg} as well as quantum diodes for the rectification of the spin~\cite{Casati2018} and thermal~\cite{Landi-diod,Casati2019} currents.
However, the possibility to switch different quantum NESS only by means of a local  magnetic field is, to the best of our knowledge, a completely uninvestigated physical phenomenon of potential interest for applications.

{\it Model.}
Let us consider the Heisenberg $XXZ$ spin chain of length $N$ coupled to dissipative magnetization baths which tend to polarize the first and last spin of the chain in the states $\psi_L\equiv\ket{\th_L,\vfi_L}$ and $\psi_R\equiv\ket{\th_R,\vfi_R}$, respectively, along the corresponding polarization vectors: $\vec{n}_\alpha \equiv (\langle \si_{i_\alpha}^x \rangle,\langle \si_{i_\alpha}^y \rangle,\langle \si_{i_\alpha}^z \rangle)$ with $\alpha=L,R$ and $i_L=1, i_R=N$.
Here  $\ket{\th_i,\vfi_i} = \ket{\cos\frac{\th_i}{2}e^{-i \frac{\vfi_i}2},\sin\frac{\th_i}{2}e^{i \frac{\vfi_i}2}}$  denotes the  pure single spin state at site $i$, $\vec \si$ are the usual Pauli matrices and $\th_i,\vfi_i$ are spherical coordinates of the local magnetization, i.e. $(\langle \si_i^x \rangle,\langle \si_i^y \rangle,\langle \si_i^z \rangle)=(\sin \th_i \cos \vfi_i,\sin \th_i \sin \vfi_i, \cos \th_i)$. The interaction with the dissipative environment is described in the framework of the Lindblad master equation (LME)~\cite{Petruccione,ClarkPriorMPA2010}:
\begin{align}
  \frac{\partial \rho(t) }{\partial t} &= {{\cal L}\left[\rho(t)\right]=} -\frac{i}{\hbar}
  \left[H_{0},\rho(t)\right] + \Gamma \sum_{i=1}^2 {\cal D}_{L_{i}} [\rho(t)],
  \label{LME}
\end{align}
where ${\cal L}$ is  the Liouvillean operator  generating the full evolution of the state and $H_0$ is the $XXZ$ spin chain Hamiltonian with uniaxial anisotropy describing the coherent part of the evolution:
\begin{align}
H_{0}&=J
\sum_{n=1}^{N-1} \left(\si^x_{n}\si^x_{n+1}+ \si^y_{n}\si^y_{n+1}+\Delta (\si^z_{n}\si^z_{n+1}-I)\right).
\label{Ham}
\end{align}
In Eq. (\ref{LME}) $\Ga$ denotes the dissipation strength, $\mathcal{D}_L X \equiv  L X L^\dagger - \frac{1}{2} (L^\dagger L X + X L^\dagger L)$, with $L_i$ Lindblad operators targeting the polarizations $\vec{n}_L$,$\vec{n}_R$ at the left (site $1$) and  right (site $N$) edges, respectively, taken as
\begin{equation}
L_1=  \ket{\th_L,\vfi_L}\bra{\th_L,\vfi_L}^\perp,\;\;\;
L_2= \ket{\th_R,\vfi_R}\bra{\th_R,\vfi_R}^\perp,
\end{equation}
with $\ket{\th,\vfi}^\perp$ denoting the state orthogonal to $\ket{\th,\vfi}$.
It is a generic feature of Eq. (\ref{LME}) that, independently on initial conditions,  after a relaxation time a NESS is reached: $\rho (t)\rightarrow\ro_{NESS}$.
Due to the gradient of magnetization created by the dissipative baths, a generic NESS is intrinsically a nonequilibrium state, unique for our choice of dissipators.
By changing the bath polarizations  $\vec{n}_L$, $\vec{n}_R$ and adjusting the anisotropy parameter, $\Delta$, one can manipulate the NESS  to obtain a pure state $\rho=\ket{SHS}\bra{SHS}$ ~\cite{2016PopkovPresilla},~\cite{2017JPAPopkovPresillaSchmidt}, with
\begin{align}
  \ket{SHS} &\equiv \prod_{k=0}^{N-1} \otimes
  \ket{\th,k\ga}=
  \prod_{k=0}^{N-1}\otimes \left(
    \begin{array}{c}
      \cos(\frac{\theta}{2})  e^{-\frac{i}{2}k \ga}
      \\
      \sin(\frac{\theta}{2})   e^{\frac{i}{2}k \ga}
    \end{array}
  \right)
  \label{SHS}
\end{align}
and $\ga=\arccos\De$. Note  that the azimuthal angle in (\ref{SHS}) increases by $\ga$ at each site along the chain so that the spins build up a fully polarized frozen helix of period $2\pi/\ga$, also called spin-helix state (SHS). To relax on a SHS the dissipation $\Ga$ must be much stronger than the coherent part of the evolution and polarization baths must fit the SHS, i.e.  $\psi_L\equiv \ket{\th,0}$, $\psi_R\equiv \ket{\th,\ga(N-1)}$. Also note that the relation  $\De=\cos\ga$ implies that SHS  with opposite chiralities, i.e. $\ket{SHS(\th,\pm\ga)}\equiv\ket{\pm}_{1,N}$, can exist for the same value of  $\Delta$ and they carry opposite  ballistic currents:
\begin{align}
_{1,N}\bra{\pm} \hat{j}_{k}^z\ket{\pm}_{1,N}&=\pm 2 J \sin \ga \sin \theta,
\label{meancurrent}
\end{align}
with $\hat{j}_{k}^z=2J(\si_k^x\si_{k+1}^y-\si_k^y\si_{k+1}^x)$ denoting the $z-$component of the magnetization current operator.

{\it Pure kink-SHS.}
In order to manipulate the spin-helix NESS by means of local magnetic fields we consider the Hamiltonian $H = H_{0}+\vec{h}_M\cdot\vec{\si}_M$ with $H_{0}$ given by (\ref{Ham}) and $\vec{h}_M$ denoting a magnetic field applied to a bulk spin at site $M$, $1<M<N$. In the following  we assume, for simplicity, boundary bath polarizations in the $XY$-plane ($\th=\pi/2$) and denote $\ket{\vfi}\equiv \ket{\pi/2,\vfi}$. The above SHS properties allow to construct a novel type of NESS consisting of two SHS pieces of opposite chirality  joined  at the site $M$. Indeed, by properly choosing $\vec{h}_M$ and the boundary bath polarizations, it is possible to stabilize via dissipation the pure NESS:
\begin{align}
\ket{+,-}_M&\equiv\ket{+}_{1,M}\otimes \ket{-}_{M+1,N}= \nonumber \\ = &\prod_{k=0}^{M-1} \otimes \ket{k\ga }\prod_{k=1}^{N-M} \otimes \ket{ (M-1-k)\ga}
\label{NESS-kink}
\end{align}
that is exact in the limit $\Ga \rightarrow \infty$ (see Supplementary Material \cite{supp}).
Obviously, such a state can exist only due to the presence of the magnetic field that sustains the change of chirality (kink) created at the site $M$. Indeed, note that at the site $M$ the state (\ref{NESS-kink}) undergoes an inversion of chirality, i.e.  $\ket{+,-}_M\equiv\prod_{k=1}^{N} \ket{\vfi_k}$ with $\vfi_{k+1}-\vfi_{k}=\ga$ for $k<M$ and  $\vfi_{k+1}-\vfi_{k}=-\ga$ for $k\geq M$.
In the following we refer to this state as a kink-spin helix state (kink-SHS).

The fact that the $\ket{+}$ and  $\ket{-}$ pieces of the kink-SHS~(\ref{NESS-kink}) carry opposite ballistic currents (see Eq. (\ref{meancurrent})) implies that the site $M$  must be either a source or a sink, depending on the sign of $\ga$, and  in order to have stationarity the following balance condition for $\langle \si_M^z \rangle$ must be satisfied:
\begin{align}
0=\frac{\partial}{\partial t} \langle \si_M^z \rangle &=
 \langle \hat{j}_{M-1}^z - \hat{j}_{M}^z -i [\sum_{\al=1}^3 h_M^\al \si_M^\al,  \si_M^z ]  \rangle.
  \label{CondBalanceJz}
\end{align}
\begin{figure}[ptbh]
\vskip-1.6cm
\begin{center}
{\includegraphics[width=0.5\textwidth]{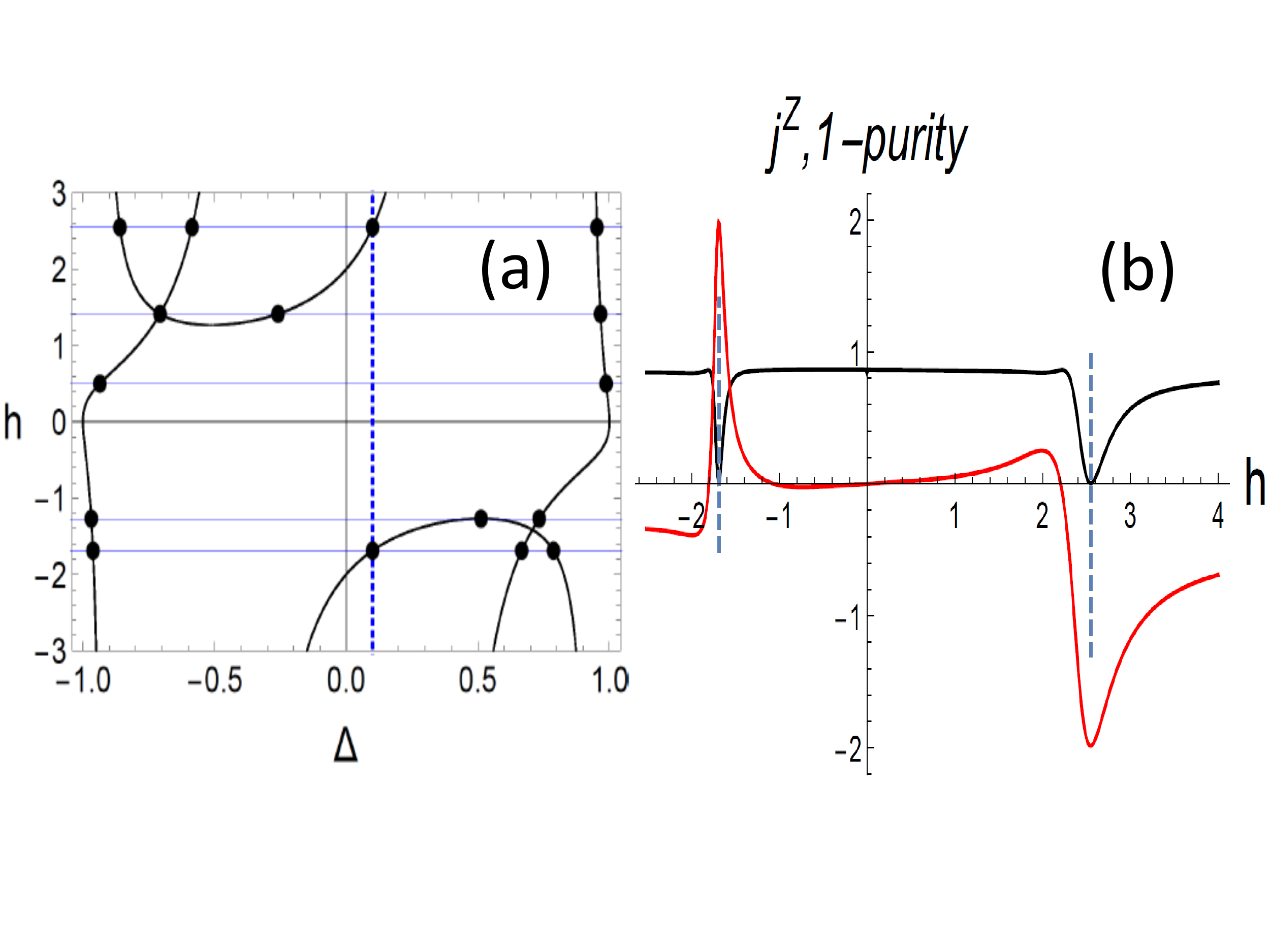}}
\end{center}
\vskip-1.8cm
\caption{Panel {\bf(a)}. Kink SHS existence region (thick continuous curves) in the parameter space $(\Delta, h)$ as obtained from Eq.~(\ref{ConstraintMagneticField}) for a spin chain
of size  $N=5$ with $J=1$ and  magnetic field $h(\sigma^x_3+\sigma^y_3)$ applied at site $M=3$. Note the symmetry of the curves under the transformation $h(\Delta)\rightarrow -h(-\Delta)$. Dots refer to kink-SHS obtained from numerical solutions of the stationary LME for $\Ga=10^3$ and  magnetic field amplitudes corresponding to the depicted horizontal thin lines $h=2.54797, 1.414215, 0.5, -1.267582, -1.688785$. Panel {\bf(b)}. Rescaled purity, $1-p$, (black curve) and steady magnetization current measured at the link between sites $1$ and $2$ of the chain (red curve) obtained from numerical solutions of the stationary LME as functions of the magnetic field amplitude $h$. Parameter values are the same as in panel (a) but $\De=0.1, \vfi_L=\vfi_R=0$. Dashed lines show the location of the pure kink-SHS $\ket{+,-}$ and $\ket{-,+}$, which are exact mirror brothers predicted by (\ref{ConstraintMagneticField}) for $h_M^x=h_M^y\equiv h$, $\ga=\pm \arccos \De$, and corresponding  to the bottom and top dots on the vertical dotted line $\De=0.1$ of the left panel, respectively.
}
\label{Fig2}
\end{figure}
Calculating the averages in (\ref{CondBalanceJz})  with respect to  the kink-SHS (\ref{NESS-kink}) and using (\ref{meancurrent}) we find:
\begin{align}
&h_M^y \cos(\ga(M-1))  - h_M^x \sin(\ga(M-1))=2 J \sin \ga.  \label{ConstraintMagneticField}
 \end{align}
Note that this condition on the magnetic field  does not involve the $h_M^z$ component. A more elaborate analysis (see Supplementary Material \cite{supp}) yields that $h_M^z=0$, together with two other conditions: i) the relation $\De=\cos \ga$ and ii) the compatibility of  the state (\ref{NESS-kink}) with the  boundary polarizations, i.e. $\psi_L =\ket{\vfi_L}, \ \  {\psi_R}=\ket{\vfi_R}$, with $\vfi_L=0$ and  $\vfi_R=(2M-N-1)\ga$.

Condition (\ref{ConstraintMagneticField}) plays a central role in the following. Indeed, for any given $N,M$, Eq.~(\ref{ConstraintMagneticField}) and the above condition i) define a hyperplane in the parameter space $\{\Delta, h_M^x, h_M^y\}$ where pure kink-SHS exist. A $2D$ cut, along the $h_M^x= h_M^y$ direction of the hyperplane is shown in Fig.~\ref{Fig2}a. By tuning the local magnetic field components in the $XY $plane according to  Eq.~(\ref{ConstraintMagneticField}) and increasing the dissipation rate $\Gamma$, the NESS can be made arbitrarily close to the pure NESS (\ref{NESS-kink}). This is shown in Fig.~\ref{Fig2}b where the  magnetization current and $1-p$, with $p=tr(\rho_{NESS}^2)$ denoting the purity, are reported  as functions of the magnetic field amplitude $h$, for the same spin chain and parameter values considered in Fig.~\ref{Fig2}a, but for $\De=0.1$ and $\vfi_L=\vfi_R=0$. From this we see that $1-p = 0$ (i.e. kink-SHS are pure states) exactly in correspondence of the predicted values of $h$ (compare with $h$ values of points on the vertical dashed line in Fig.~\ref{Fig2}a). Moreover, the current at these points attains  the same  maximum amplitude but with opposite signs, as  expected for opposite chirality, $\ket{+,-}$, $\ket{-,+}$ kink-SHS. Also note that since the kink-SHS condition $\De=\cos\ga$ couples the anisotropy to the boundary magnetization gradient $\vfi_R-\vfi_L$, a change of the anisotropy or a change of the chirality (i.e. $\ga\rightarrow -\ga$), makes necessary an adjustment of the magnetization gradient. If $N$ is odd and the field is applied at the center of the chain, however,  the constraint  $\vfi_R=(2M-N-1)\ga$ is satisfied for aligned borders, $\vfi_L = \vfi_R=0$, for arbitrary $\ga$.
\begin{figure}[tbp]
\vskip-1.6cm
\begin{center}
{\includegraphics[width=0.48\textwidth]{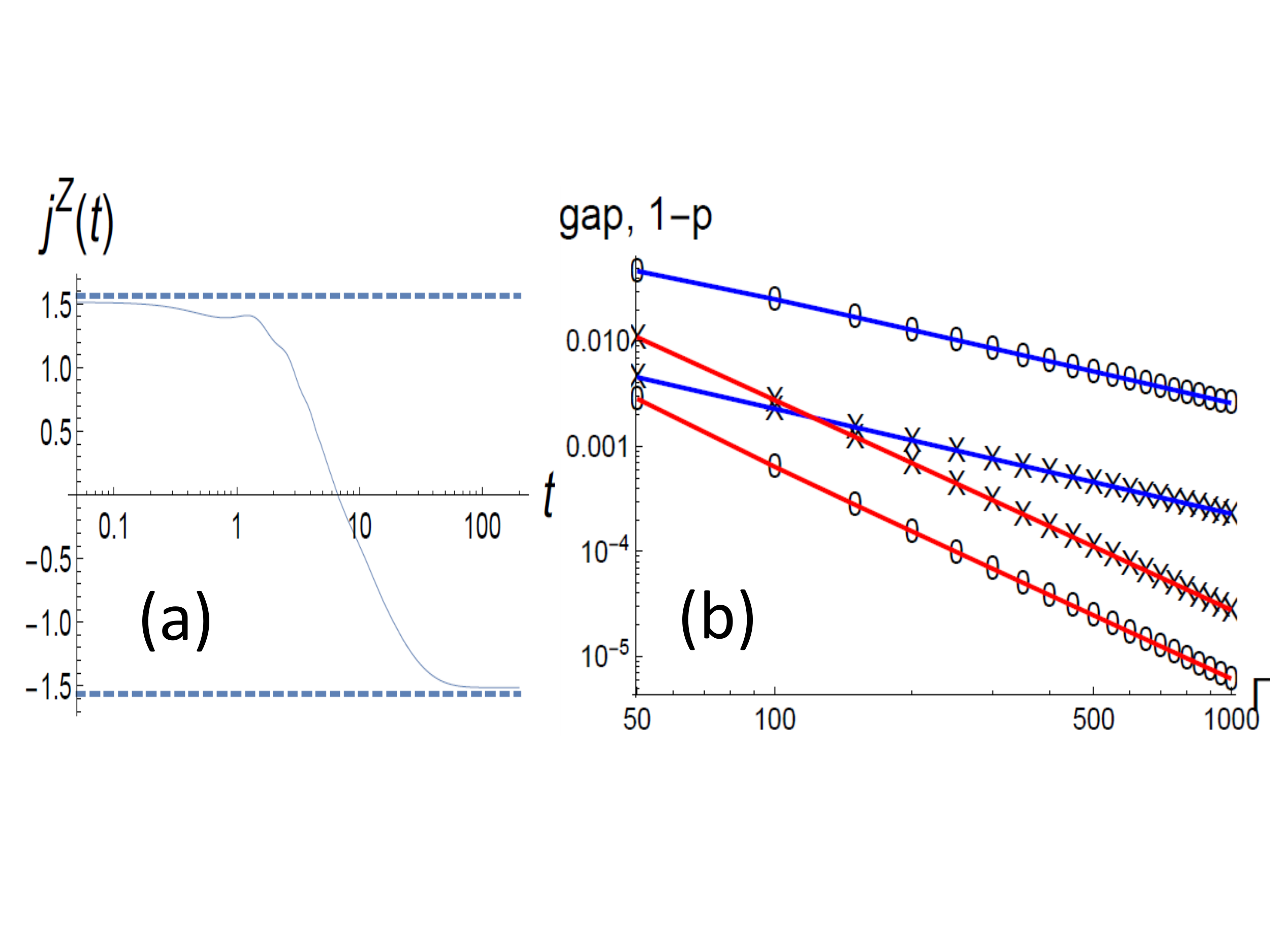}}
\end{center}
\vskip-1.6cm
\caption{Panel {\bf(a)}.Time evolution of the magnetization current $j^z$ measured at the link between sites $1,2$ after an abrupt change of a magnetic field, $(h_M^x,h_M^y,0) \rightarrow (h_M^x, -h_M^y,0)$ at time $t=0$, applied to the middle site. Initial condition is a right-handed kink-SHS (the NESS for $h_M^y<0$), final state is the left handed kink-SHS (the NESS for $h_M^y>0$). Parameters values are $J=1, N=5, M=3$, $\Ga=20$, $\Delta= \cos 2\pi/7, \vfi_L=\vfi_R=0$. Local field amplitudes
$h_M^x=-1.775, h_M^y=0.75$, correspond to the optimal choice indicated by the dotted line in
Fig.~\ref{Fig4} and satisfy the constraint (\ref{ConstraintMagneticField-alignedBorders}) with $"+"$ sign. Dashed lines show the magnetization currents $\langle j^z \rangle=2 \sin(2\pi/7)$ of the asymptotic Zeno states. Panel {\bf(b)}. Numerical data for the Liouvillean gap (upper data points joined by blue lines) and for $1-tr \rho^2$ (lower data points joined by red lines) of the NESS as function of $\Gamma$, illustrating the scaling properties in (\ref{DefGap}).
Parameter values are fixed  as in panel (a) except for markers "$\times$" for which $h_M^x=-2.0603, h_M^y=2.0$. Blue and red lines with slopes $1/x$ and $1/x^2$, respectively, are drawn to guide the eyes.
}
\label{Fig3}
\end{figure}
%

{\it A kink-SHS quantum switch.}
 Let us consider a chain of odd size $N$ with a local magnetic field  applied to the central site, $M=(N+1)/2$  and  $\vfi_R=\vfi_L=0$. In this case the bath reservoirs are the same for both edges (as sketched in Fig.~\ref{fig-Switch})) and Eq.~(\ref{ConstraintMagneticField}) is written as:
\begin{align}
   h_M^y \cos\frac {(N-1)\ga}{2} \mp h_M^x \sin\frac {(N-1)\ga}{2}&=\pm 2 J\sin \ga,
 \label{ConstraintMagneticField-alignedBorders}
 \end{align}
where the different signs correspond to the kink-SHS $\ket{+,-}$ and $\ket{-,+}$ obtained from Eq.~(\ref{NESS-kink}) with $+\ga$ and $-\ga$, respectively.

From Eq. (\ref{ConstraintMagneticField-alignedBorders}) it follows that, by keeping  the magnetic field component amplitude $h_M^y$ fixed  and flipping its sign, one can  pass from the pure state $\ket{+,-}$ to the pure state $\ket{-,+}$ (or viceversa), realizing in this manner a quantum dissipative switch that operates only via the local magnetic field.
This is illustrated in Fig.~\ref{Fig3}a where the time evolution  of the magnetization current $j^z$, obtained from direct numerical integrations
of the LME (\ref{LME}), is shown after an instantaneous change of the magnetic field. In the numerical simulation the initial condition is taken as a  kink-SHS of type $\ket{+,-}$, created  by tuning the local field $(h_M^x,h_M^y,0)$ to the  corresponding anisotropy value using Eq.~(\ref{ConstraintMagneticField-alignedBorders}),
and by flipping, at the time $t=0$, the $y$-component of the field. The net outcome of the flip is the change  of the initial state $\ket{+,-}$ into the kink-SHS $\ket{-,+}$ of opposite chirality, this giving the inversion of the current shown in Fig.~\ref{Fig3}a.
\begin{figure}[tbp]
\centerline{
\includegraphics[width=6.8cm,height=5.7cm,clip]{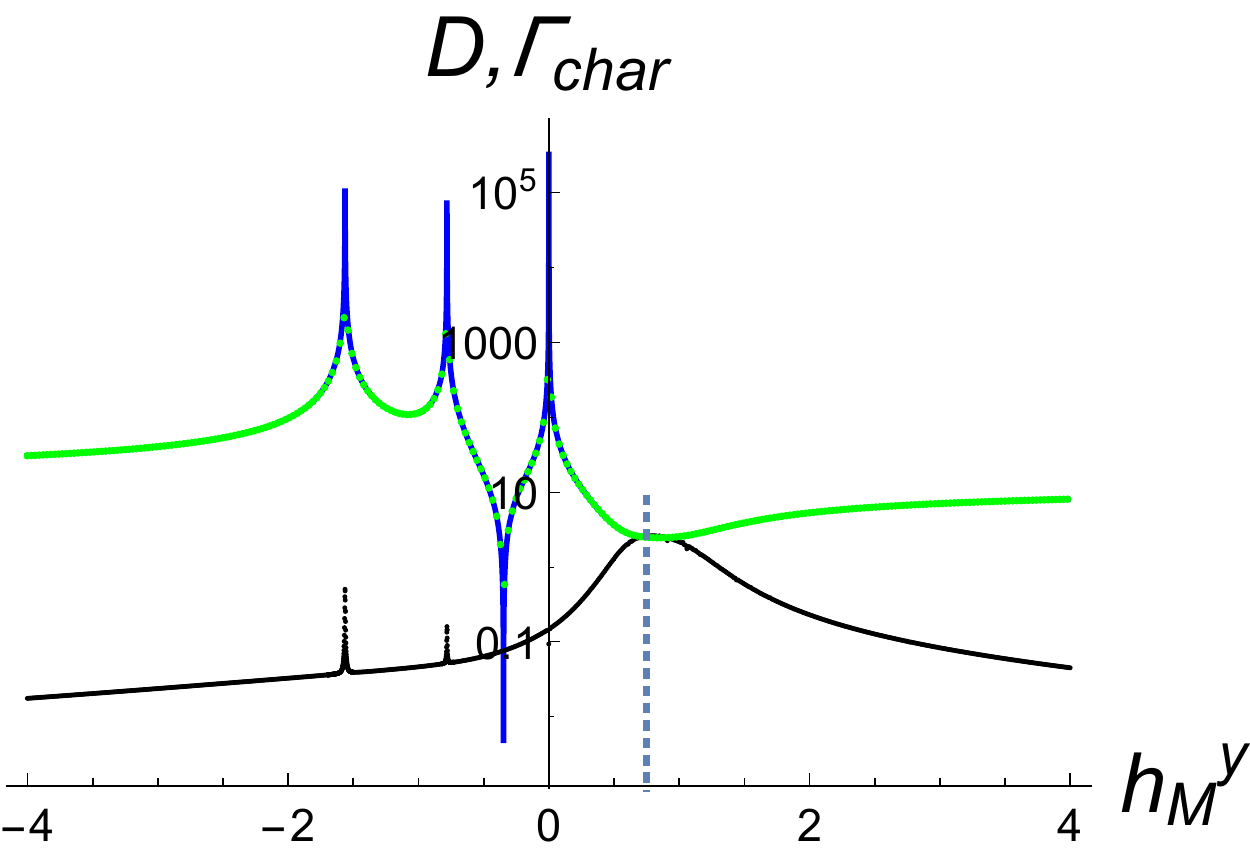}
}
\caption{$\Gamma_{char}$ (upper dotted green line) and $D$ (lower dotted black line) versus $h_M^y$ as obtained  from direct numerical calculations of Eq.~(\ref{LME})  for parameter values fixed as in Fig.~\ref{Fig3}a but with $\Gamma=3000$. The continuous blue line refers to exact analytical expression of $\Ga_{char}$ derived in Eq. $(9)$ of the Supplementary Material~\cite{supp}. The optimal choice of $h_M^y \approx 0.75$ is indicated by the dotted vertical line.
}
\label{Fig4}
\end{figure}
%

{\it Fidelity and relaxation time optimization.}
Although exact pure kink-SHS are attained for  infinitely large dissipations, for any practical purpose it is possible to achieve an  effective Zeno regime when the  dissipation becomes larger than some  characteristic finite value, $\Gamma_{char}$,  which depends on system parameters.
For finite $\Gamma$ the distance of the NESS from  the asymptotic Zeno pure state can be measured with the  purity $p$ while  the relaxation time, $\tau$, i.e. the time  needed to reach the asymptotic state, is of the order $\tau \approx 1/g$ where $g$ is the gap of the Liouvillean superoperator associated to the LME (\ref{LME}). In Fig.~\ref{Fig3}b we show the Liouvillean gap and the  purity obtained from direct numerical solutions of the LME, as functions of $\Ga$, from which we see that:
\begin{align}
1-p&=\frac{\Gamma_{char}^2}{\Gamma^2}+ o(\Gamma^{-2}),\;\;\;\;\;\;
g=\frac{D}{\Gamma}+ o(\Gamma^{-1}),
\label{DefGap}
\end{align}
with $\Ga_{char}$, $D$, depending on system parameters but not on $\Ga$.
From this it follows that for a better and faster convergence to the Zeno pure NESS one must minimize $\Gamma_{char}$ and maximize $D$. This can be done by observing that Eq.~(\ref{ConstraintMagneticField-alignedBorders}) implies the existence of a one-parameter family of solutions for $h_M^x,h_M^y$ which corresponds exactly the the same kink-SHS, this allowing to choose one of the components of the magnetic field freely. Taking  the $y$-component $h_M^y$ as a free parameter,  we have investigated in Fig.~\ref{Fig4} how $D$ and $\Gamma_{char}$ depend on it. We see that the dependence on $h_M^y$ is huge and, for the considered case, the optimal choice is obtained in the interval $h_M^y \approx 0.75 \pm 0.25$, where $D$ and $\Ga_{char}$ reach their max and min, respectively (note that $D$ and $\Gamma_{char}$ may vary by orders of magnitude outside this interval). Moreover, for opposite  chirality $\pm \ga_0$ states we find
\begin{align}
D(h_M^y,+\ga_0)&= D(-h_M^y,-\ga_0),\label{Prop-Dsymmetry}
\\
\Gamma_{char}(h_M^y,+\ga_0)&= \Gamma_{char}(-h_M^y,-\ga_0),\label{Prop-Gammasymmetry}
\end{align}
these being a consequence of the mirror symmetry of the respective physical states and imply that the $D, \Ga_{char}$ optimization holds true for both $(\pm h_M^y, \pm \ga_0)$ parameter choices, i.e. for both kink-SHS involved in the switch.

{\it Conclusion.} We have demonstrated the possibility to use local magnetic fields to switch between different  pure NESS, preserving their purity and  attaining optimal spin current. A quantum switch that inverts the current via the local field only, was discussed for the case of "aligned borders". The developed theory, however, is general and results can be extended to arbitrary bath polarizations. The optimization of the relaxation time and fidelity of the NESS involved in the switch was also discussed. We believe that the above properties and  the robustness of kink-SHS against decoherence makes these NESS of interest  for applications.

{\it Acknowledgements.}
V.P. acknowledges support from the  European Research Council (ERC) through the advanced grant 694544 -- OMNES and from the DFG grant KL 645/20-1. M.S. acknowledges support from the Ministero dell'Istruzione, dell'Universit\'a e della Ricerca (MIUR) through the grant PRIN-2015-K7KK8L on "Statistical Mechanics and Complexity".


\end{document}